# Mapping axonal density and average diameter using non-monotonic time-dependent gradient-echo MRI

Daniel Nunes[1], Tomás L Cruz[1], Sune N Jespersen[2,3], Noam Shemesh*[1]

[1] Champalimaud Neuroscience Programme, Champalimaud Centre for the Unknown, Av. Brasilia 1400-038, Lisbon, Portugal
[2] Center of Functionally Integrative Neuroscience (CFIN) and MINDLab, Clinical Institute, Aarhus University, Aarhus, Denmark.
[3] Department of Physics and Astronomy, Aarhus University, Aarhus, Denmark

*Corresponding Author
Noam Shemesh
Champalimaud Neuroscience Programme, Champalimaud Centre for the Unknown
Av. Brasilia 1400-038
Lisbon, Portugal
E-mail: noam.shemesh@neuro.fchampalimaud.org
Phone number: +351 210 480 000 ext. #4467

**Running title:** Multi-gradient Echo imaging of axonal density and diameter

**Abbreviations**

CNS – Central Nervous System; dCST – dorsal corticospinal tract; DTI – Diffusion Tensor Imaging; EPI – Echo Planar Imaging; FG – Fasiculus Gracilis; FC – Fasiculus Cuneatis; FOV – Field of View; GLM – General Linear Model; GLTA – Generalized Lorenzian Tensor Approach; MGE – Multi-Gradient-Echo; MRI – Magnetic Resonance Imaging; NMR – Nuclear Magnetic Resonance; OGSE – Oscillating Gradient Spin-Echo; PBS – Phosphate Buffer Saline; QSI – q-space Imaging; ReST – Reticulospinal tract; RST – Rubrospinal tract; SNR – signal to noise ratio; SAR – Specific Absorption Rate; STT – Spinothalamic tract; TE – echo time; TR – repetition time; ; VST – Vestibulospinal tract; WM – White Matter.

**Highlights:**

- Axonal density and average size are critical neurophysiological factors targeted for MRI
- A simple microstructural model for Multi-Gradient Echo (MGE) MRI is suggested
- Axon density-dependent signal non-monotonicities are predicted for MGE
- Experiments conducted on rat spinal cords at 16.4 T validate the predictions
- Microstructural parameters extracted from MGE correlate with histology
- Suggests potential of microstructural mapping using the SNR-efficient MGE sequence

**Graphical Abstract**

## Multi-Gradient-Echo signal oscillations...

Ex-vivo rat spinal cord

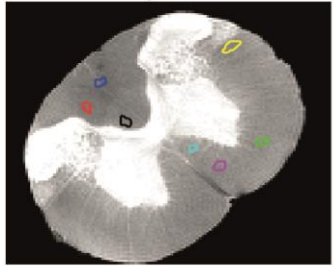
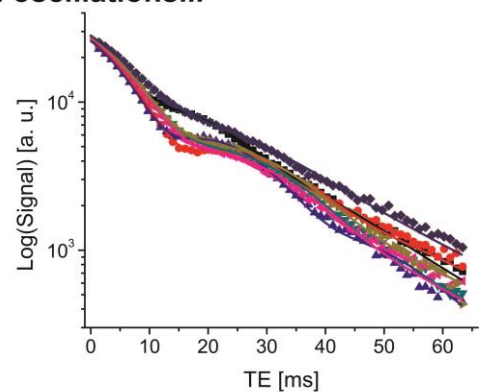

## ...mapping of axon diameter and density

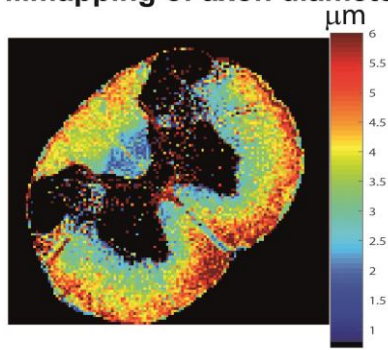
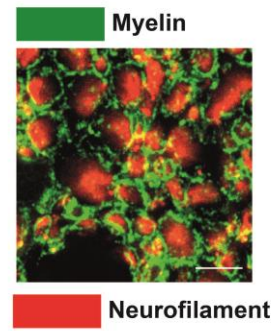

Myelin
Neurofilament

# Abstract

White Matter (WM) microstructures, such as axonal density and average diameter, are crucial to the normal function of the Central Nervous System (CNS) as they are closely related with axonal conduction velocities. Conversely, disruptions of these microstructural features may result in severe neurological deficits, suggesting that their noninvasive mapping could be an important step towards diagnosing and following pathophysiology. Whereas diffusion based MRI methods have been proposed to map these features, they typically entail the application of power gradients, which are rarely available in the clinic, or extremely long acquisition schemes to extract information from parameter-intensive models. In this study, we suggest that simple and time-efficient multi-gradient-echo (MGE) MRI can be used to extract the axon density from susceptibility-driven non-monotonic decay in the time-dependent signal. We show, both theoretically and with simulations, that a non-monotonic signal decay will occur for multi-compartmental microstructures – such as axons and extra-axonal spaces, which we here used in a simple model for the microstructure – and that, for axons parallel to the main magnetic field, the axonal density can be extracted. We then experimentally demonstrate that maps derived from MGE acquired at 16.4 T in *ex-vivo* spinal cords, where the different tracts characterized by different microstructures are clearly contrasted in parametric maps extracted by fitting the MGE decay to the model. When the quantitative results are compared against ground-truth histology, they seem to reflect the axonal fraction (though with a bias, as evident from Bland-Altman analysis). As well, the extra-axonal fraction can be estimated. The results suggest that our model is oversimplified, yet at the same time evidencing a potential and usefulness of the approach to map underlying microstructures using a simple and time-efficient MRI sequence. We further show that a simple general-linear-model can predict the average axonal diameters from the four model parameters, and map these average axonal diameters in the spinal cords. While clearly further modelling and theoretical developments are necessary, we conclude that salient WM microstructural features can be extracted from these simple, SNR-efficient multi-gradient echo MRI, and that this paves the way towards easier estimation of WM microstructure *in vivo*.

# Introduction

Axon density, degree of myelination, and the regional size distribution, all play a paramount role in both healthy and diseased Central Nervous System (CNS) function (1). In normal white matter (WM), axonal conduction velocities are determined by these properties, with larger and more myelinated axons producing higher conduction velocities (2-6), which in turn facilitate rapid information transduction between remote CNS regions (2,4,7-9). Severe deficits arise from even slight aberrations to the axonal microstructure: changes in the axonal size distribution and myelin structure in optic nerve lead to severe and progressive vision loss (10,11); decreases in axon density (axonal loss) is observed in Multiple-Sclerosis (MS) histology (12), and areas characterized by a specific range of axonal diameters may be more prone to axonal loss (12,13). Axonal losses, as well as changes in average size and myelination, can be found in the vicinity of MS plaques but also in normal appearing white matter (14), suggesting their involvement in slow and potentially chronic damage to the tissue. Upon trauma, dramatic changes in average axon size, density, and myelination are observed with time (15,16). Other microstructural abnormalities, such as axonal beading, are thought to be involved in, e.g., stroke (17). Such changes to the neuronal morphology can be followed by pronounced cognitive impairments (18).

Several Magnetic-Resonance-Imaging (MRI) methods have been proposed for mapping axonal microstructures, and in particular, diffusion weighted imaging was shown to map regional average axon size changes in diseased tissues (19-24). In normal CNS, Ong et al have demonstrated that the histologically well-characterized spatial distribution of mean axon diameter in the spinal cord can be mapped noninvasively by q-space imaging (QSI) using a gradient system capable of producing 50 T/m (25,26). Assaf et al proposed the AxCaliber model for characterizing variations in the (histologically well-known (27)) axon diameter distributions in the corpus callosum (28) (though note recent concerns (29)), and Duval et al used AxCaliber on the 300 mT/m gradients of the Connectome Scanner to map spinal cord axonal diameters in-vivo in humans (30). Very recently, Xu et al demonstrated that axon microstructure in the rat spinal cord could be estimated (31) using oscillating gradients spin echo (OGSE) diffusion MRI, showing good correlations between the OGSE-derived maps and histology (31). Non-uniform oscillating gradients MRI – a technique exhibiting greater sensitivity towards smaller dimensions (32) – was recently shown to contrast the corpus callosum's histologically known (27) five different tracts (33). When powerful gradients are not available (34) the axon index can be mapped (35), which preserves some of the axon diameter contrast (36). Other diffusion-based methods use more sophisticated modelling (35,37-43) or pulse sequences (44-49) to extract other features of white matter, such as its underlying orientation dispersion, neurite density, and microscopic anisotropy.

Susceptibility-driven contrasts (50,51) have been gaining increasing attention in recent years, both in terms of anatomical contrasts, and, more recently, in terms of microstructure. Lee et al showed that phase images arising from gradient-echo data in white matter show a strong orientation dependence with respect to the main magnetic field (52); Liu introduced susceptibility tensor imaging (STI), which provides information on the absolute orientation of anisotropic systems when samples are rotated with respect to the main magnetic field (53) or when a more elaborate multipole scheme is employed (54). Several models have been put forward to describe the biophysical origins of the susceptibility anisotropy (55-66). $T_2^*$ anisotropy has likewise been emerging as a highly useful contrast for orientation-mapping in WM (67-71), and several models have again been put forward to explain its origins (72-74). Non-mono-exponential signal decay has been recently observed in WM when gradient echo measurements reached relatively high TEs, suggesting the contributions of multiple microstructural compartments to the signal, and allowing their spatial mapping (73,75). Chen et al recently simulated how multiple compartments would impact the full signal decay, finding a pattern in orientation and TE-dependences of the signal (76), which, importantly, could potentially be used to map specific compartments within the white matter. Importantly, the new contrasts revealed in these studies were derived from one of MRI's simplest sequences: the MGE, which is time-efficient and typically high in SNR. However, to our knowledge, MGE's potential to actually extract – or even qualitatively contrast – crucial microstructural metrics such as axonal density and average sizes has not been heretofore studied, though an analysis of non-exponential relaxation in the context of microstructural disorder has been presented very recently (77).

In this study, we harness a simplistic model of white matter tissue and its full susceptibility-driven time-dependent signal response, to establish that the axon density can in fact be determined from simple MGE experiments. We then validate our theoretical findings in *ex-vivo* rat spinal cords, demonstrating experimentally that remarkable axon density contrasts can be obtained that highlight the major tracks within the SC, and which correlate with histological findings. We further show that other parameters of the model, such as the susceptibility-driven frequency shift, seem to qualitatively reflect the regional variation in average axon diameters. The potential of time-dependent MGE experiments for characterizing more specific features in WM are discussed.

## Theory

An object placed in a homogeneous magnetic field **B₀**, will impart a local shift in the Larmor frequency $\Delta\omega(\mathbf{r})$ of a magnitude proportional to the susceptibility difference $\Delta\chi$ between water and the object. In general, there is a complex relationship between $\Delta\omega(\mathbf{r})$ and $\Delta\chi$ (55-57,61-66) depending on the objects' shape and geometrical arrangement. Here we restrict ourselves to a very simple model of spinal cord white matter as a collection of parallel cylinders representing the axons oriented along the main magnetic field. In this case, the induced magnetic fields inside and outside axons are homogeneous but differ by an amount proportional to $\Delta\chi$. Furthermore, echo times are assumed to be sufficiently large to ignore the signal contribution from myelin water, as we hypothesized that at the ultrahigh field used for these experiments (16.4 T), myelin water is likely to have very short $T_2^*$. Hence, the magnitude signal can be computed from a sum of two terms: an intra-cylindrical compartment with a volume fraction $f_i$, arbitrarily selected to be on-resonance, and an extra-cylindrical compartment with a volume fraction $1 - f_i$ and frequency shift $\Delta\omega$. Each compartment is assumed to exhibit monoexponential transverse relaxation with relaxation rates $R_{2i} = 1/T_{2i}$ and $R_{2e} = 1/T_{2e}$, corresponding to intra- and extra-axonal relaxation rates, respectively. Hence, the magnitude signal can be written as:

$$S(TE) = S_0 \left| f_i e^{-\frac{TE}{T_{2i}^*}} + (1-f_i) e^{-TE(\frac{1}{T_{2e}^*} + i\Delta\omega)} \right| = S_0 *$$

$$\sqrt{f_i^2 e^{-2TE*R_{2i}^*} + 2f_i(1-f_i) * e^{-TE(R_{2i}^* + R_{2e}^*)} * \cos(\Delta\omega * TE) + (1-f_i)^2 e^{-2TE*R_{2e}^*}}$$

Eq. 1

Equation 1, though extremely simple, is central in this study, because it predicts a non-monotonic and oscillatory signal decay, from which the four model parameters can be extracted. This obviates the need to extract the signal phases in every TE, which may involve quite elaborate signal processing, yet still obtain the information contained within the model.

In reality, we expect small deviations from axonal misalignment with the main magnetic field, which will induce inhomogeneous magnetic field in the compartments. Violations of axons as simple cylinders will have similar implications, and such effects will lead to a more complicated time-dependent phase shift between the compartments, as well as non-exponential $T_2^*$ decay in each of the compartments. Nevertheless, here we assume these deviations to be sufficiently small and hence restrict ourselves to exploring the potential of the simple model in Eq. 1 for characterizing spinal cord microstructure.

# Methods

*Simulations.*

Susceptibility-driven signals were simulated using in-house code written in MatLab (The Mathworks, Natick, MA, USA). We assume that any given white matter voxel is composed of identically oriented (though not necessarily parallel to the field), perfectly cylindrical axons, and that water fills intra- and extra-axonal spaces. The simulations are performed in the static dephasing regime, i.e., diffusion effects were not considered. To simulate the mesh of axons we started by automatically generating randomly distributed non-overlapping circles in a 100x100$\mu m^2$ square. The desired values of axonal diameter (in this case, d=3 μm) was set, and the target density (in our case, varying between 0.015-0.4 for different simulations) was obtained by iteratively removing or generating new circles, until the targeted density / diameter values were reached. Finally, a binary mask was generated if a voxel was designated as being intra-cylindrical.

To compute the magnetic susceptibility within each voxel, the points inside the axons were attributed a bulk susceptibility value of -80 ppb and to the points outside the axons a bulk susceptibility value of 0 ppb. To simulate the magnetic field arising from these susceptibility distributions, the binary susceptibility distribution generated before was convolved with the dipole field kernel (60), an expression which, for $B_0$ aligned cylinders only, will coincide with the more accurate and comprehensive Generalized Lorenzian Tensor Approach (GLTA) (64):

$$\Delta\omega(x,y,z) = \gamma B_0 \mathcal{F}^{-1}\left\{\left(\frac{1}{3} - \frac{k_z^2}{k^2}\right) \cdot \mathcal{F}\{\chi(x,y,z)\}\right\} \quad \text{Eq. 2}$$

where $\mathcal{F}$ denotes a Fourier transform, k the wavenumber, and χ the susceptibility's spatial distribution. To avoid the boundary effects caused by the finite simulation, only the central part of the voxel (50x50x50$\mu m^3$) was used for the estimation of the NMR signal. This was then plugged into Eq. 1 to simulate a MGE signal at 16.4T assuming the cylinders were perfectly parallel to the field. For the simulations used in this work, we chose $T_{2i}^* = 25\ ms$ and $T_{2e}^* = 10\ ms$ (these values were selected based on our experimental results). For Figure 9, fields of {3,7}T were simulated with $T_{2i}^* = \{80, 50\}\ ms$ and $T_{2e}^* = \{30, 10\}\ ms$, respectively. In the context of this study, only magnitude data were considered, and the magnitude signal decays were plotted as function of increasing TEs.

*Specimen preparation.*

All experiments were preapproved by the Champalimaud Centre for the Unknown's ethics committee. Male Long Evans rats (*N=7*) were perfused transcardially with 4% paraformaldehyde, followed by isolation and extraction of the spinal cord. The seven extracted spinal cords were stored in a 4% PFA solution for at least four days prior to any NMR experiments. Before undergoing imaging, the spinal cords were washed with PBS and placed in a fresh PBS solution for at least 12 h. Cervical spinal cord sections were cut into ~5 cm pieces and placed with their long axis parallel to the orientation of the main magnetic field in a 5 mm NMR tube, which was subsequently filled with Fluorinert (Sigma Aldrich, Lisbon, Portugal). A smaller, Fluorinert-filled NMR tube served as a plunger to prevent the spinal cord from floating upwards.

*MRI experiments.*

All experiments were performed on a 16.4 T Bruker Aeon scanner, interfaced with an Avance IIIHD console and equipped with a gradient system capable of producing up to 3000 mT/m in all directions. The following single slice scans were performed on the 7 spinal cords:

*Diffusion Tensor Imaging.* DTI-EPI experiments were performed to test the alignment of spinal cords using the following parameters: FOV=4.8 x 3.8 $(mm)^2$, matrix size of 132x96, in-plane resolution of 34x34 $(\mu m)^2$ (a partial Fourier encoding was used with 50% coverage), and a slice thickness of 500 µm, as in the MGE experiments, and TR/TE = 7000 / 28.5 ms. The diffusion parameters were set to Δ/δ = 12/2.5 ms and a b-value of 1.2 ms/µm$^2$, and 15 orientations were acquired, as well as 6 non-diffusion weighted images for normalization.

*Multi-Gradient Echo experiments.* Once it has been established that the spinal cords are parallel to the field, multi-gradient echo experiments were conducted using the following parameters: FOV = 4.8x3.8 $(mm)^2$, matrix size of 132x96 (no partial Fourier encoding), leading to an in-plane resolution of 34x34 $(\mu m)^2$, and a slice thickness of 500 µm. The MGE sequence was run using a TR/flip angle combination of 200 ms / 26°, and with an acquisition bandwidth of 156250 Hz which facilitated minimal TE spacing. TEs were varied between {1.400:1.106:65.654} ms (60 increments, all echoes were used, with no discernable phase artifacts between positive/negative echoes), and 1200 averages were acquired, leading to a total experimental duration of ~7.5 h. Another identical MGE scan was acquired but with only 380 averages, leading to an experimental time of 2 h.

### *Histology.*

Immunohistochemistry (IHC). Upon MGE imaging, the spinal cords were embedded in 2% agarose for sectioning using a vibratome (Leica, Germany). Free floating horizontal sections 50 μm thick were collected from the imaged region. Afterwards, the spinal cord sections were incubated in vehicle, for blocking and permeabilization, containing 5% normal goat serum (Sigma-Aldrich, cat.# G9023), 1% bovine serum albumin fraction V (Sigma-Aldrich, cat.# A9418), 0.3% Triton X-100 (Sigma-Aldrich, cat.# T9284) in PBS pH=7.4, during 30 min. Next, the sections were incubated overnight, at 4ºC under gentle agitation, with the primary antibodies anti-myelin proteolipid protein (PLP; Novus Biologicals, cat.# NB100-1608) and anti-neurofilament 160/200 (NF; Sigma-Aldrich, cat.# N2912) at 5 μg/ml diluted in vehicle solution. Subsequently, the sections were washed in vehicle three times, 10 minutes each, and incubated with the secondary antibodies Alexa Fluor dye conjugated (Alexa Fluor 647 donkey anti-mouse IgG, and Alexa Fluor 488 goat anti-chicken IgG; Thermofisher, Cat.# A31571 and Cat.#A11039, respectively) for 1 h, at room temperature, and light protected. Then, the sections were washed in PBS with 0.3% Trition X-100, followed by a 20 min incubation with 500 ng/ml of DAPI (Sigma-Adrich, cat.# D9542) in PBS with 0.3% Trition X-100, at room temperature, light protected. Afterwards, the sections were washed twice in PBS with 0.3% Trition X-100, during 10 min each and then mounted in glass coverslips with self-made moviol. The coverslips were tightened with transparent nail polisher to avoid water losses, in order to avoid changes in the refractive index of the mounting medium.

*Microscopy.* A Zeiss LSM 710 laser scanning confocal microscope was used (Zeiss, Germany) for immunohistochemistry image acquisition. To acquire the full spinal cord area, a 6x4 tile scan using a 10X objective (EC Plan Neofluar, numerical aperture = 0.3, Zeiss, Germany) was made in confocal mode. Afterwards, each ROI was imaged using a 63X immersion objective (Plan Apochromat, numerical aperture = 1.4, Zeiss, Germany) in confocal mode. To avoid cross-talk between absorption and emission spectra, the scanning was performed sequentially using the 405 nm, 488 nm and 633 nm lasers. Immersion oil with refractive index of 1.43, similar to moviol was used. Images were acquired without using of averaging filters. Low resolution images had a voxel size of (X,Y,Z) of 0.83x0.83x0.3 μm$^3$ with X-Y matrix of 1024x1024 pixels. Image acquisition was made at 200 Hz to achieve a pixel dwell time of about 6.3 μs. High resolution images had a voxel size of (X,Y,Z) of 70x70x150 nm$^3$ with X-Y matrix of 2048x2048 pixels. Image acquisition was made at 100 Hz to achieve a pixel dwell time of 6.3 μs.

To ensure that the slices acquired in the microscope were from as close a location as possible to the MRI images analyzed in the study, we first acquired low resolution images on the confocal microscope to have an overview of the slice. Next, much higher resolution images were acquired in the coordinates corresponding to the ROIs used for the MRI image analysis (measured from specific landmarks in the MR images that could also be seen in the low resolution microscopy image). This ensured that the ROIs placed in noninvasive MRI and invasive histology were as identical as possible.

*Image analysis.*

*Diffusion Tensor Imaging.* The diffusion tensor was reconstructed by custom-written code in MatLab using nonlinear fitting to the data. A white matter mask was manually selected, and the angle between $B_0$ and the eigenvector corresponding to the primary eigenvalue was calculated.

*Multi-Gradient Echo experiments.* Data arising from the MGE experiments was first examined via ROI analysis. Seven ROIs were placed within the major fiber tracts as previously performed *(25,26,31)*, and the mean signal value within the ROI was computed for each TE. These data were fit using Matlab's powerful lsqcurvefit function, to Eq. 1. Eq. 1 has five unknowns, $S_0$, $f_i$, $R_{2i}^*$, $R_{2e}^*$ and $\Delta\omega$ which were extracted for each ROI from the fitting. The initial parameters for lsqcurvefit were [max(S(TE)) [a.u.], 0.5, 1/70 ms$^{-1}$, 1/120 ms$^{-1}$, 0.25 rad/sec], respectively, for each of the 5 unknowns; extensive preliminary testing found that the fitting quality does not strongly depend on the initial parameters, and that the fit converges to the same values even when the initial parameters vary. Importantly, since a set of parameters $f_i, R_{2i}^*, R_{2e}^*$ and a different set of parameters $(1-f_i), R_{2e}^*, R_{2i}^*$, will give an equally good fit to the data, we always assigned the smaller $R_2^*$ component to the cylindrical compartment. Following the ROI analyses, we extracted the full parametric maps for the five unknowns by a pixel-by-pixel fitting of the experimental data to Eq. 1.

*Immunohistochemistry data analysis.* Images were processed using the ImageJ software (free software and plugins for image processing and analysis in Java; accessible at imagej.nih.gov/ij/). Confocal single frames were processed using a 2D anisotropic diffusion filter for noise removal. Additionally, noise from pixels with high intensity were removed by Fourier transform of the spatial data in its frequency domain where a 5 pixels radius mask was applied at theta=0° and r=134.95 µm/cycle, to filter noise with high frequencies. Afterwards, using the inverse Fourier transform algorithm filtered space domain images were obtained. Background suppression and image smoothing were performed, keeping edges unaltered, by applying a Fourier transform bandpass filter, with a high frequency cut-off corresponding to 40 pixels in image space and low spatial frequency cut-off corresponding to 5 pixels in image space. This method for image processing is more efficient than the conventional median filters and convolution algorithms *(78)*. Segmentation of the image frames was obtained by making it binary, in order to calculate the histological axonal fraction $h_a$ and the histological internal fraction, $h_{int}$ corresponding to the fraction of the area occupied by myelin+axons, respectively, $h_{int} = h_a + h_m$. To do this, the "analyze particles" algorithm implemented in ImageJ was used. All statistical analysis were performed using Matlab.

Bland-Altman analysis was performed by plotting the difference of MRI- and IHC-driven metrics against their mean, and plotting the data along with the 95% confidence interval. Since the sample size was 49 (smaller than the 60 recommended for using z statistics), we used the more stringent definition of the confidence intervals, $CI = x \pm t_{0.05, n-1} s \sqrt{1 + \frac{1}{n}}$ where x is the mean difference, s is the standard deviation, n=49, and t indicates the t-statistic.

*General Linear Model (GLM) analysis.* To investigate a putative (and assumed) linear relationship between the MRI data and the histological ground-truth, we used Matlab's fitlme (fit linear mixed effects) function to regress the MRI data onto the histological findings. Despite that our model (Eq. 1) precludes a dependence on axon diameter, we empirically observed quite clear relationships, and following preliminary tests showing that no specific parameter clearly correlates, we attempted to use the general linear model with all four parameters. Hence, we attempted to characterize these phenomenologically. In particular, the mean axon diameter observed in histology, $h_d$ was assumed to be of the form

$$h_d = c_0 + c_1 f_i + c_2 T_{2i}^* + c_3 T_{2e}^* + c_4 \Delta\omega$$

Eq. 3

where the coefficients $c_i$ correspond to the linear regression coefficients. The parameters were calculated from all spinal cords used in this study.

# Results

Simulations.

Figure 1 shows simulations for MGE sequences for different cylinder density (here, representing the axonal density). Since the orientation of the cylinders was parallel to the field, the frequency maps (Fig. 1A) show only two specific frequencies: namely, a zero, on-resonance frequency (blue) and an offset frequency (yellow, corresponding to $\gamma\Delta\chi B_0$). As the density increases, the proportion of each frequency within a given volume varies accordingly, but the frequencies remain constant. For very low axon density, the signal will decay in the expected nearly mono-exponential fashion (Fig. 1A); however, as the axon density increases, the proportion of offset spins increases, resulting in a signal non-monotonicity which is readily observed in magnitude data (Fig. 1B). Figure 1C simulates the TE-dependent signal decay, but for a somewhat narrower cylindrical fraction range of 0.25-0.35, which reflects more realistic axon density variations in normal CNS. The signal dip appears clearly in these simulations, and variation can be seen with respect to signal minima and shape for each of the different densities, despite that they are quite closely spaced. Hence, the signal carries at least some signature towards the axon density. Importantly, the signal oscillations predicted by Figure 1 are axon radius independent and hence – in this ideal scenario – they will only reflect the axon density.

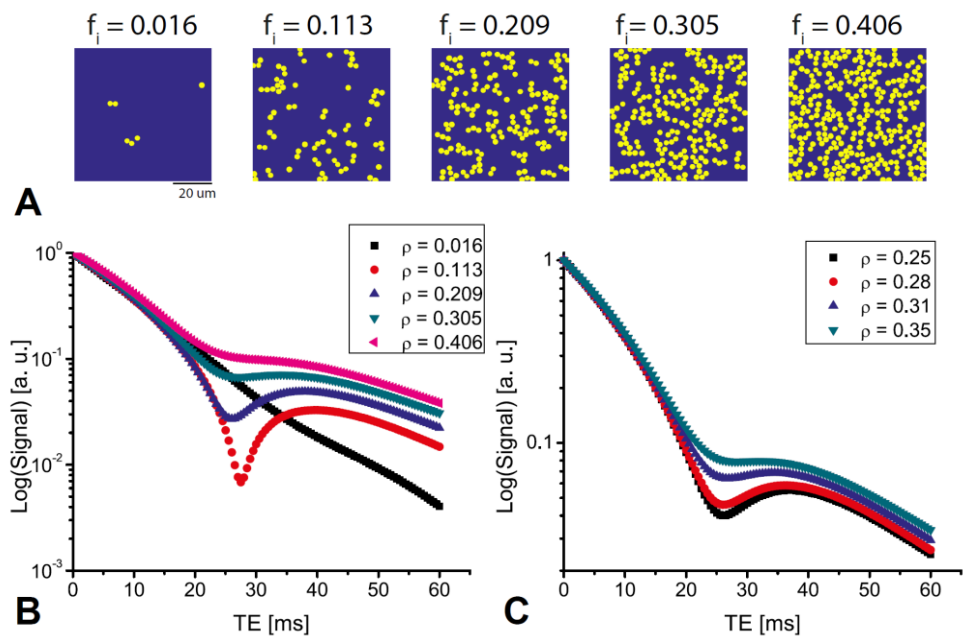

**Figure 1.** Simulations for MGE signal decays in perfectly aligned cylinders. (A) Frequency distribution maps for the microstructures considered here. Cylinder density was varied from ~0.1-0.4. Notice that two discrete frequencies emerge in these plots (blue and yellow). (B) Echo-time dependence of the magnitude of the signal, $|S(TE)|$. Notice the density-dependent signal behaviour, including the non-monotonic, diffraction-like behaviour. $f_i$ denotes the intra-cylindrical fraction. (C) Further simulations conducted for the narrower range $f_i$ = 0.25-0.35, showing that densities have quite unique signatures for the microstructure in these MGE magnitude data.

MRI experiments.

Prior to execution of the MGE experiments, it is imperative to ensure that the spinal cords are indeed as parallel as possible with respect to $B_0$. Figure 2 (upper panel) shows the result of the DTI scans, in particular, a representative color-coded fractional anisotropy map, with blue denoting the z-direction (parallel to $B_0$). The spinal cord in general, and the WM regions in particular were found to be very well aligned with their principal axis pointing along the direction of the main magnetic field. A histogram of the polar angle distribution in an ROI placed within the WM is shown in Fig. 2 (lower panel), revealing that the vast majority of pixels are aligned within +/- 3° (though note this does not address potential intravoxel orientation dispersion), with the histogram maximum centered very close to zero. Similar trends were observed across all spinal cords used in this study.

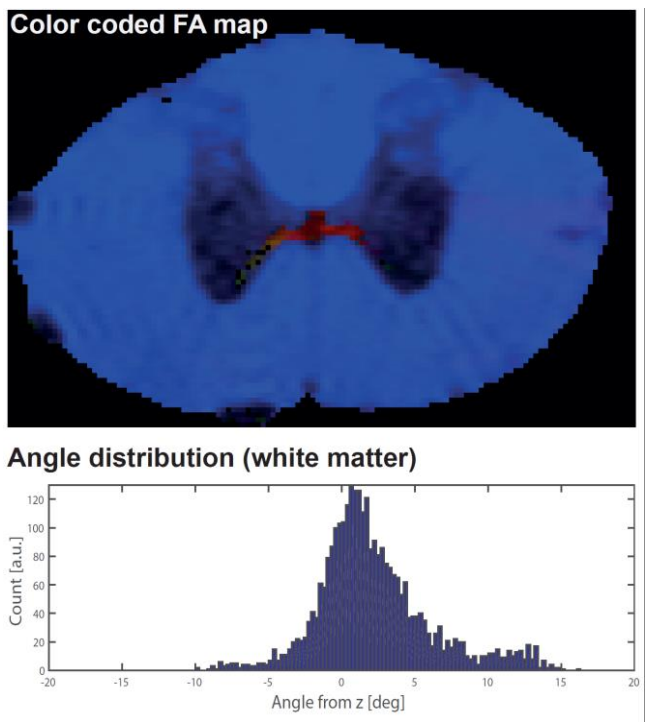

**Figure 2.** Alignment of the spinal cord with respect to B0 in a representative rat spinal cord. Color-coded FA maps extracted from the DTI scan reveal predominantly parallel orientation in white matter (blue denotes the magnet's z-direction, the direction of the field). Red color reflects left-right orientation and green up-down. The histogram of angles in the entire white matter in the slice shows that the majority of white matter voxels are aligned within ~+/-3° of the z-axis.

Having established that the spinal cords are well aligned with the field, we sought to experimentally validate the occurrence of the signal oscillations predicted in Fig. 1. Figure 3 shows raw data from three representative spinal cords (symbols) as well as fits to the model (solid lines), for the seven major WM tracts within the spinal cord. A signal non-monotonicity was observed in most ROIs, throughout the different spinal cords. Some ROIs, such as those placed within the dCST (black symbols), exhibited a less pronounced oscillatory nature of the signal decay. These non-exponential, and in most cases, non-monotonic signal decays were observed in all spinal cords studied using the MGE sequence (N=7). The typical signal to noise ratio in the first image was ~200, though the same non-monotonic dependence was also observed for much lower initial SNR. Table 1 summarizes the main fitted parameters arising from the data within these seven spinal cords, namely, the fraction of internal spins, the respective $T_{2i}^*$ and $T_{2e}^*$ and the frequency shift between the compartments. The extracted parameters exhibit high reproducibility, with the highest standard error (SE) < 2 % for the internal fractions, < 8 % for $T_{2i}^*$ and < 5 % for both $T_{2e}^*$ and $\Delta\omega$.

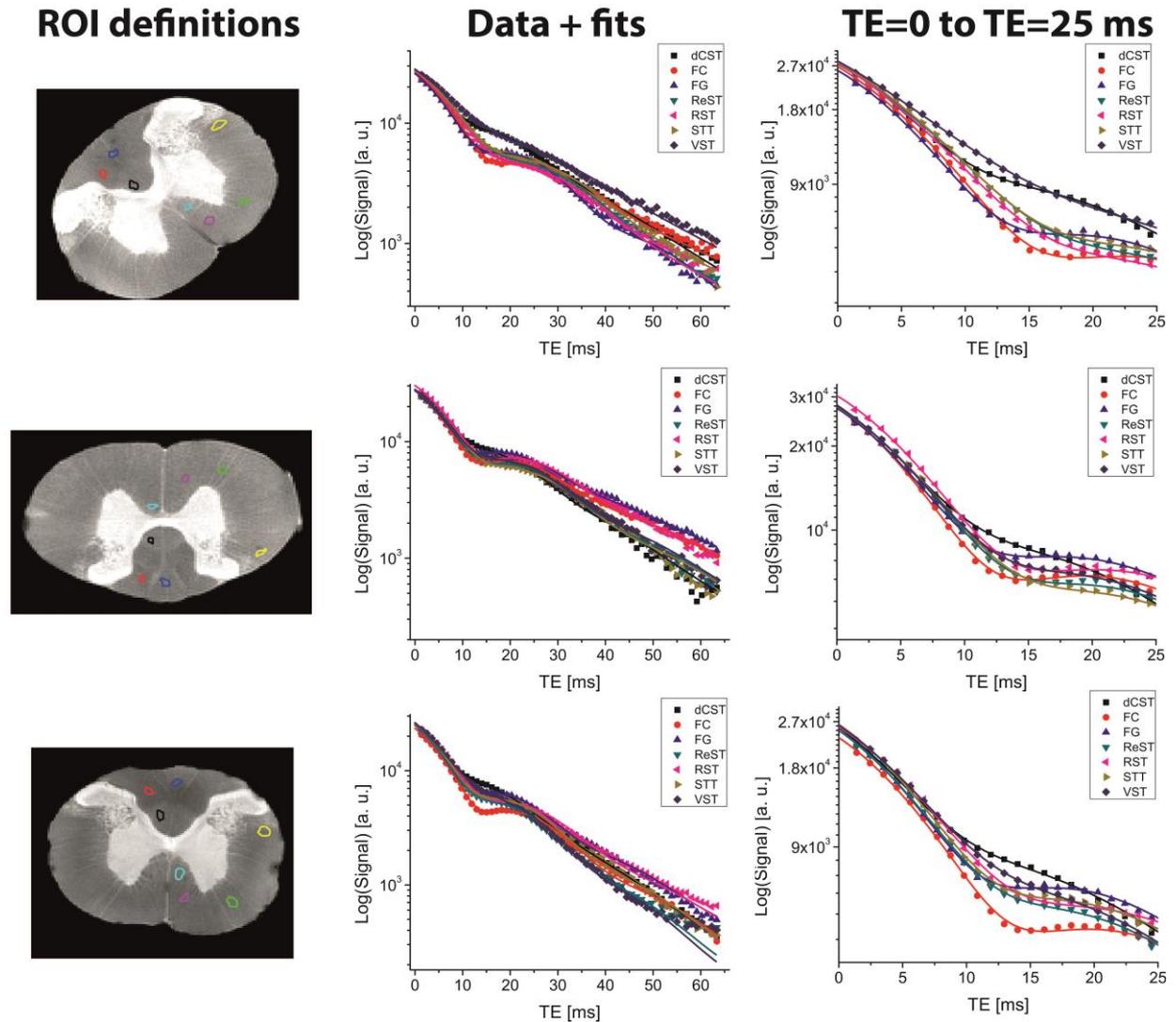

**Figure 3.** ROI analysis in three spinal cords. ROIs were placed within the seven major tracts running along the spinal cord (left column). The average signal in the ROI is plotted in the middle column, showing the signal decay (symbols) and the fits (solid lines) to the data using Eq. 1. A consistent non-monotonicity was observed for numerous ROIs, in a consistent fashion across specimens. The rightmost column expands the TE=0-25 ms region, to better highlight the oscillations and fits. The signals are plotted in log scale for easier viewing of the oscillation. dCST – dorsal corticospinal tract; FG – Fasiculus Gracilis; FC – Fasiculus Cuneatis; ReST – Reticulospinal tract; RST – Rubrospinal tract; STT – Spinothalamic tract; VST – Vestibulospinal tract.

Fits to 3-compartment models did not demonstrate significant improvements over the two compartment fit (not shown).

**Table 1**. Model parameters extracted from the MGE MRI experiments for the different ROIs.

| Quantity / Region | dCST | FG | FC | ReST | RST | STT | VST |
|---|---|---|---|---|---|---|---|
| Internal fraction ($f_i$) | | | | | | | |
| **Mean** | 0.79 | 0.61 | 0.66 | 0.68 | 0.65 | 0.68 | 0.76 |
| **Std. Deviation** | 0.03 | 0.03 | 0.06 | 0.04 | 0.03 | 0.02 | 0.02 |
| **Std. Error** | 0.01 | 0.01 | 0.02 | 0.02 | 0.01 | 0.01 | 0.01 |
| **Lower 95% CI of mean** | 0.76 | 0.57 | 0.61 | 0.64 | 0.62 | 0.66 | 0.74 |
| **Upper 95% CI of mean** | 0.81 | 0.64 | 0.72 | 0.72 | 0.68 | 0.70 | 0.78 |
| **Minimum** | 0.76 | 0.56 | 0.57 | 0.61 | 0.61 | 0.64 | 0.73 |
| **Maximum** | 0.85 | 0.65 | 0.73 | 0.74 | 0.72 | 0.70 | 0.78 |
| $T_{2i}^*$ [ms] | | | | | | | |
| **Mean** | 17.00 | 19.86 | 19.31 | 17.91 | 18.91 | 17.74 | 17.57 |
| **Std. Deviation** | 1.67 | 2.82 | 4.01 | 1.72 | 2.62 | 1.44 | 1.79 |
| **Std. Error** | 0.63 | 1.07 | 1.51 | 0.65 | 0.99 | 0.55 | 0.68 |
| **Lower 95% CI of mean** | 15.45 | 17.25 | 15.60 | 16.32 | 16.49 | 16.40 | 15.91 |
| **Upper 95% CI of mean** | 18.55 | 22.46 | 23.01 | 19.51 | 21.34 | 19.07 | 19.23 |
| **Minimum** | 14.59 | 15.78 | 14.68 | 15.72 | 15.99 | 16.14 | 14.45 |
| **Maximum** | 19.78 | 23.78 | 25.23 | 20.75 | 23.30 | 19.98 | 19.88 |
| $T_{2e}^*$ [ms] | | | | | | | |
| **Mean** | 7.69 | 9.35 | 9.17 | 9.89 | 9.49 | 9.92 | 9.34 |
| **Std. Deviation** | 0.47 | 0.43 | 1.02 | 1.04 | 0.94 | 0.85 | 0.72 |
| **Std. Error** | 0.19 | 0.17 | 0.42 | 0.42 | 0.38 | 0.35 | 0.29 |
| **Lower 95% CI of mean** | 7.20 | 8.90 | 8.10 | 8.80 | 8.50 | 9.03 | 8.58 |
| **Upper 95% CI of mean** | 8.18 | 9.79 | 10.25 | 10.98 | 10.47 | 10.81 | 10.09 |
| **Minimum** | 7.20 | 8.94 | 8.23 | 8.12 | 8.25 | 8.84 | 8.27 |
| **Maximum** | 8.22 | 10.03 | 11.17 | 11.09 | 10.66 | 11.01 | 10.44 |
| $\Delta\omega$ [Hz] | | | | | | | |
| **Mean** | 43.06 | 34.51 | 38.00 | 34.67 | 34.54 | 34.62 | 35.22 |
| **Std. Deviation** | 1.66 | 2.67 | 2.11 | 3.61 | 4.04 | 3.67 | 3.72 |
| **Std. Error** | 0.68 | 1.09 | 0.86 | 1.47 | 1.65 | 1.50 | 1.52 |
| **Lower 95% CI of mean** | 41.32 | 31.72 | 35.79 | 30.88 | 30.30 | 30.78 | 31.31 |
| **Upper 95% CI of mean** | 44.80 | 37.31 | 40.21 | 38.46 | 38.77 | 38.47 | 39.12 |
| **Minimum** | 40.97 | 29.30 | 35.27 | 28.34 | 27.86 | 28.76 | 27.74 |
| **Maximum** | 44.60 | 36.14 | 40.88 | 38.50 | 39.01 | 38.03 | 37.88 |

To assess the potential of these signal oscillations for contrasting different SC regions and mapping some of its microstructural features, parametric maps were extracted from pixel-by-pixel regressions of the data to Eq. 1. Figure 4 shows parametric maps of the intra-cylindrical fraction $f_i$, intra-cylindrical relaxation constant, $T_{2i}^*$ the extra-cylindrical relaxation constant, $T_{2e}^*$, and the frequency shift between the two compartments, for three representative spinal cords.

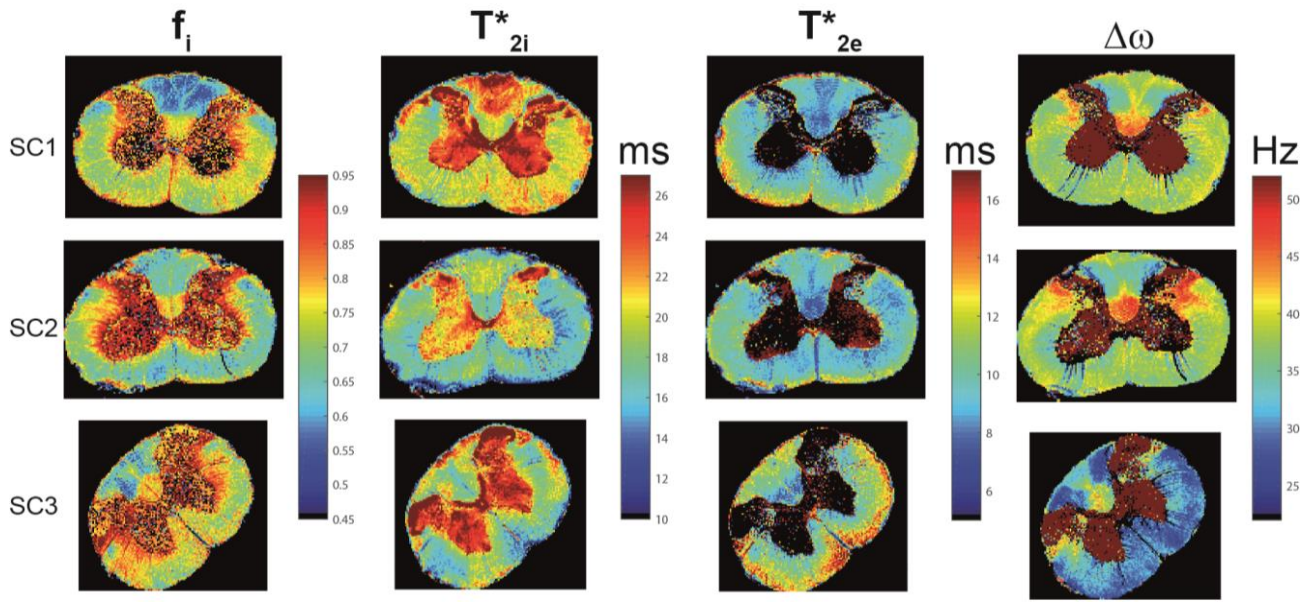

**Figure 4.** Parametric maps arising from pixel-by-pixel fits of the TE-dependence for three representative spinal cords. These maps revealed highly consistent internal fractions $f_i$, relaxation constants, and frequency shifts (SC3 represents the lower-most observed values in the entire study). A pronounced contrast between the major spinal cord tracts is observed in these maps, suggesting their correlation with underlying microstructural features.

From a qualitative perspective, these maps are clearly capable of contrasting the different and microstructurally distinct sub-regions of spinal cord WM. The fraction maps demonstrate a strong contrast between dCST and its neighbouring FC and FG tracts, as well as similarities between the fractions of dCST and VST, which, despite having very different axonal radii, are characterized by a similar axon density (cf. *(25,31)* and Tables 2 and 3). Both relaxation constant maps ($T_{2i}^*$ and $T_{2e}^*$) also show some contrast between these areas. Interestingly, the frequency shift map appears to have a somewhat different contrast compared to the fraction maps: whereas the latter shows highly different shifts between the two most microstructurally different tracts (dCST and VST), the former shows very similar axonal fractions. Furthermore, the frequency shift maps seem to qualitatively provide the most contrast between the different microstructurally distinct tracts in the WM in terms of average axon size. Finally, it is worth noting that these trends are highly consistent along these three representative spinal cords, and indeed throughout all spinal cords scanned in this study.

Table 2. Histological findings for axonal, myelin, and nuclei fractions for the different ROIs.

| Quantity/Region | dCST | FG | FC | ReST | RST | STT | VST |
|---|---|---|---|---|---|---|---|
| Axonal fraction ($h_a$) | | | | | | | |
| **Mean** | 0.35 | 0.34 | 0.30 | 0.25 | 0.30 | 0.27 | 0.32 |
| **Std. Deviation** | 0.03 | 0.03 | 0.02 | 0.03 | 0.03 | 0.03 | 0.03 |
| **Std. Error** | 0.01 | 0.01 | 0.01 | 0.01 | 0.01 | 0.01 | 0.01 |
| **Lower 95% CI of mean** | 0.32 | 0.31 | 0.29 | 0.22 | 0.27 | 0.25 | 0.30 |
| **Upper 95% CI of mean** | 0.37 | 0.36 | 0.31 | 0.28 | 0.32 | 0.29 | 0.35 |
| **Minimum** | 0.29 | 0.29 | 0.28 | 0.19 | 0.27 | 0.21 | 0.27 |
| **Maximum** | 0.37 | 0.37 | 0.32 | 0.29 | 0.33 | 0.30 | 0.37 |
| Myelin fraction ($h_m$) | | | | | | | |
| **Mean** | 0.39 | 0.38 | 0.36 | 0.35 | 0.36 | 0.36 | 0.38 |
| **Std. Deviation** | 0.02 | 0.03 | 0.01 | 0.02 | 0.03 | 0.02 | 0.02 |
| **Std. Error** | 0.01 | 0.01 | 0.008 | 0.01 | 0.01 | 0.01 | 0.01 |
| **Lower 95% CI of mean** | 0.38 | 0.36 | 0.35 | 0.33 | 0.34 | 0.35 | 0.37 |
| **Upper 95% CI of mean** | 0.41 | 0.41 | 0.37 | 0.37 | 0.38 | 0.37 | 0.40 |
| **Minimum** | 0.36 | 0.34 | 0.34 | 0.33 | 0.32 | 0.34 | 0.34 |
| **Maximum** | 0.42 | 0.42 | 0.37 | 0.38 | 0.40 | 0.39 | 0.41 |
| Nuclei fraction ($h_n$) | | | | | | | |
| **Mean** | 0.06 | 0.04 | 0.02 | 0.02 | 0.03 | 0.03 | 0.04 |
| **Std. Deviation** | 0.03 | 0.02 | 0.01 | 0.01 | 0.02 | 0.01 | 0.03 |
| **Std. Error** | 0.01 | 0.01 | 0.00 | 0.00 | 0.01 | 0.00 | 0.01 |
| **Lower 95% CI of mean** | 0.04 | 0.02 | 0.01 | 0.01 | 0.01 | 0.02 | 0.02 |
| **Upper 95% CI of mean** | 0.08 | 0.06 | 0.03 | 0.03 | 0.05 | 0.04 | 0.06 |
| **Minimum** | 0.03 | 0.01 | 0.01 | 0.01 | 0.00 | 0.02 | 0.01 |
| **Maximum** | 0.10 | 0.08 | 0.03 | 0.04 | 0.05 | 0.06 | 0.09 |

**Table 3.** Histological study of axonal diameters for the different ROIs.

| Quantity/Region | dCST | FG | FC | ReST | RST | STT | VST |
|---|---|---|---|---|---|---|---|
| Average axon diameter[1] ($h_d$) [μm] | | | | | | | |
| **Mean** | 1.43 | 2.29 | 3.39 | 4.09 | 2.66 | 2.53 | 3.43 |
| **Std. Deviation** | 0.27 | 0.26 | 0.67 | 0.49 | 0.42 | 0.59 | 0.44 |
| **Std. Error** | 0.09 | 0.09 | 0.23 | 0.17 | 0.15 | 0.21 | 0.15 |
| **Lower 95% CI of mean** | 1.20 | 2.07 | 2.83 | 3.67 | 2.30 | 2.03 | 3.06 |
| **Upper 95% CI of mean** | 1.66 | 2.51 | 3.96 | 4.50 | 3.01 | 3.033 | 3.80 |
| **Minimum** | 1.11 | 1.79 | 2.56 | 3.35 | 2.16 | 1.83 | 2.83 |
| **Maximum** | 1.95 | 2.57 | 4.31 | 4.75 | 3.47 | 3.34 | 4.24 |

[1]Inner diameter, as measured by the NF staining.

Histology.

To investigate whether the intra-cylindrical fraction maps indeed correspond to the axonal fraction in the tissue or to other microstructural elements, we stained the spinal cords using markers for axons (NF), myelin (PLP), and cell bodies (DAPI). Figure 5 shows the axon (red) and myelin (green) stains acquired from a representative spinal cord, and Table 2 summarizes the fractions of each component for the seven major spinal cord tracts. Axon density varies only slightly in the different regions, and is typically observed between 0.25-0.35, with a < 1% SE. Myelin density varies even less, between 0.35-0.4, again with a ~1% SE for all regions. Importantly, cell nuclei are small but non-negligible, showing a density of ~0.02-0.06 in the different areas. Table 3 summarizes the histological findings for mean axon sizes in these regions. As expected (25,31), the smallest axons were found in dCST, and the largest in the ReST and VST, with axons diameter varying quite highly between ~1.4 and 4.9 μm.

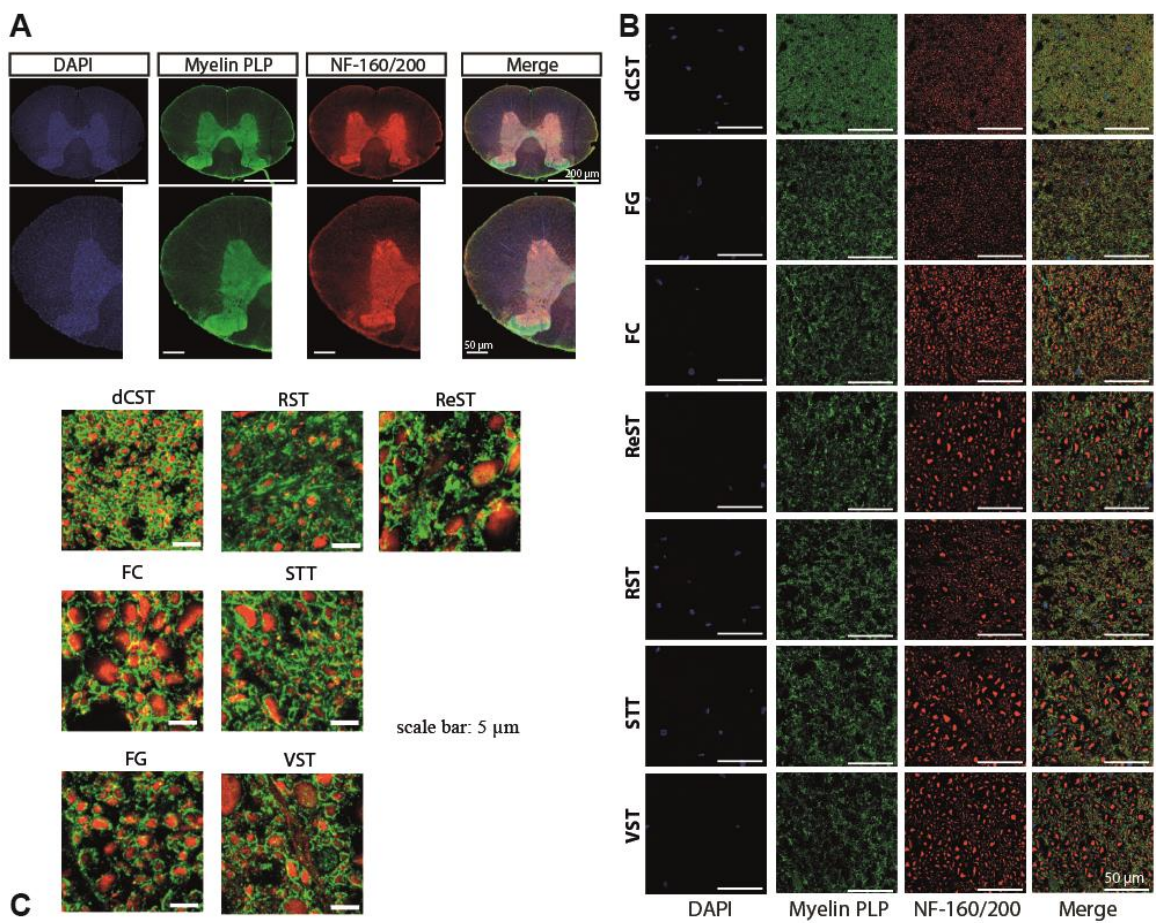

**Figure 5.** Histological study. **(A)** Wide field images of the three stains used in this study, designed to capture cell nuclei (DAPI, blue), myelin (PLP, green), and axons (NF, red). **(B)** Multichannel analysis of the different stains in different regions. Note the overlay contains no overlapping signals, showing that the antibodies were indeed selective for their target. Scale bar, 50 μm. **(C)** A zoomed view of (B) for the seven different regions of interest in the spinal cord white matter. These data were used for the quantification of all histologically-derived parameters.

These results allowed the robust quantification of the ground-truth axonal density, $h_a$, as well as the ground-truth internal fraction $h_{int}$, which takes into account both myelin and axonal density, $h_{int} = h_a + h_m$. To investigate how the parameters extracted from MRI correspond to the IHC ground-truth, Figure 6A shows the plot of the intra-cylindrical compartment obtained from the MGE scans, $f_i$, against the intra axonal fraction $h_a$ obtained from immunohistochemistry. The two parameters clearly do not correlate (Spearman's correlation coefficient, ρ=0.25, p>0.07), and interestingly, the intra-cylindrical fraction obtained from MRI is much higher compared to the axonal fractions. As one plausible explanation for this is the different normalization of IHC and MRI assuming myelin water relaxes fast, we tested the correlation of $f_i$ against $h_a/(1-h_m)$, the axonal fraction in the absence of myelin. Figure 6B shows a scatter which is closer to the unity line, but still, the correlation is not evident, nor is it statistically significant (ρ=0.275, p>0.055). We therefore compared the internal fractions with the internal water fraction obtained from IHC, i.e., with $h_{int}$ (Figure 6C). The scatter now follows the ground truth more accurately, and the Spearman correlation coefficient now reached a level of statistical significance (ρ=0.30, p<0.03). Another interesting comparison corresponds to the second fraction computed from the intra-cylindrical fraction derived from MRI, $f_e = 1 - f_i$, which can be plotted against the fraction of extra-axonal space extracted from IHC (Figure 6D). Again, a reasonably good correspondence is observed between the MRI-driven data and the extra-axonal fraction derived from IHC, $h_e$, with the extra-axonal fractions correlating with fairly high significance (ρ=0.35, p<0.015). Figure 6E shows also correlations for the myelin-relaxed extra-axonal fraction (ρ=0.30, p<0.035).

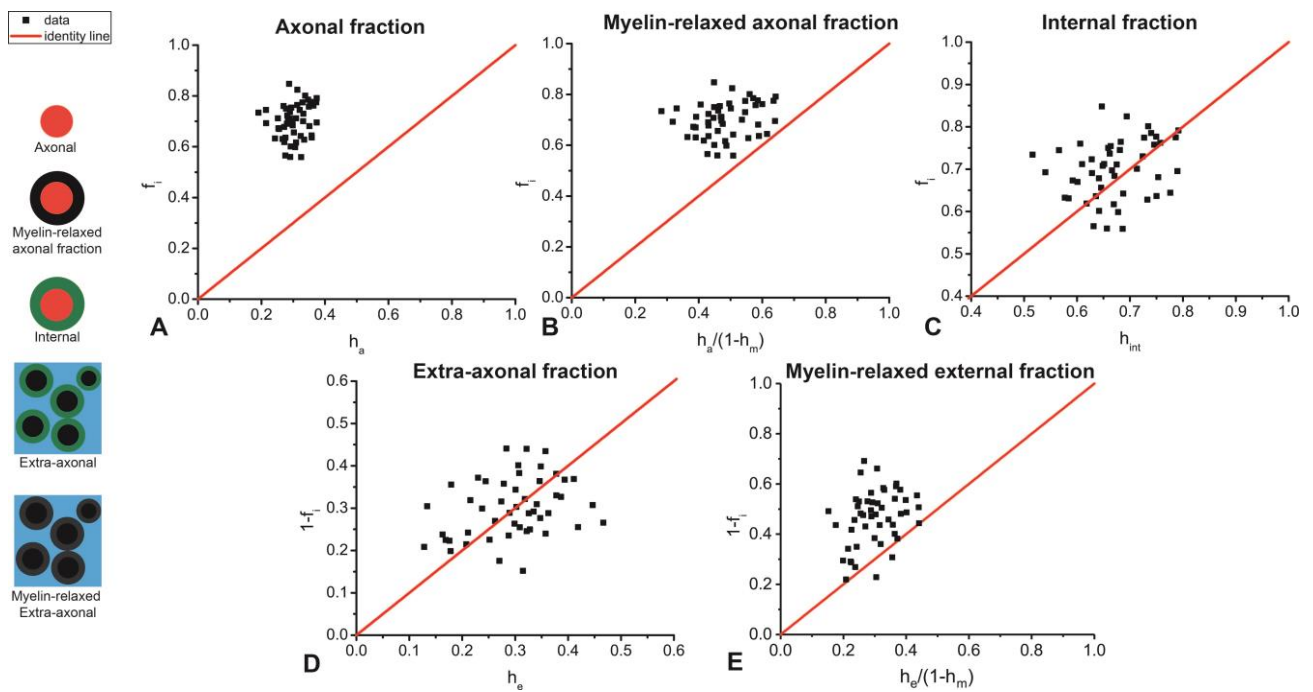

**Figure 6.** Correlation between MRI-derived parameters and IHC-derived parameters. The axonal fraction contains only intra-axonal space; the internal fraction contains both axonal and myelin spaces; and the extra-axonal water contains no water from the internal fraction. Scatter plots of the MRI-driven intra-cylindrical fraction $f_i$ against the IHC-derived **(A)** axonal fraction $h_a$, **(B)**, myelin-relaxed axonal fraction $h_a/(1-h_m)$, **(C)** internal fraction $h_{int}=h_a+h_m$. **(D)** MRI-extracted extra-cylindrical fraction compared against the extra-axonal fraction derived from IHC, **(E)** myelin-relaxed axonal fraction $h_e/(1-h_m)$. Note that the MRI-derived $f_i$ correlates well with the IHC-derived internal fraction $h_{int}$ but not with the axonal fraction $h_a$. The MRI-derived $f_e = 1-f_i$ correlates well with the histological extra-axonal fraction.

To test whether some bias may exist in Figs 6A and 6B, we performed a Bland-Altman analysis, shown in Figure 7, with Fig. 7A corresponding to scatter of $f_i$ with $h_a$, and Fig 7B and 7C corresponding to the myelin-relaxed comparison of $f_i$ against $h_a/(1-h_m)$ or $f_e$ against $h_e/(1-h_m)$, respectively. Clearly, the Bland-Altman plot shows that the pure axonal fraction is not in agreement with the intra-cylindrical. By contrast, the myelin-relaxed comparison shows that, although a bias exists, the two methods are in agreement with respect to the estimation of the axonal fraction when myelin is assumed to be "invisible" due to relaxation (Eq. 1). Figure 7C shows moderate agreement for the myelin-relaxed extra-axonal fraction, which again suggests a rather coherent bias in the data.

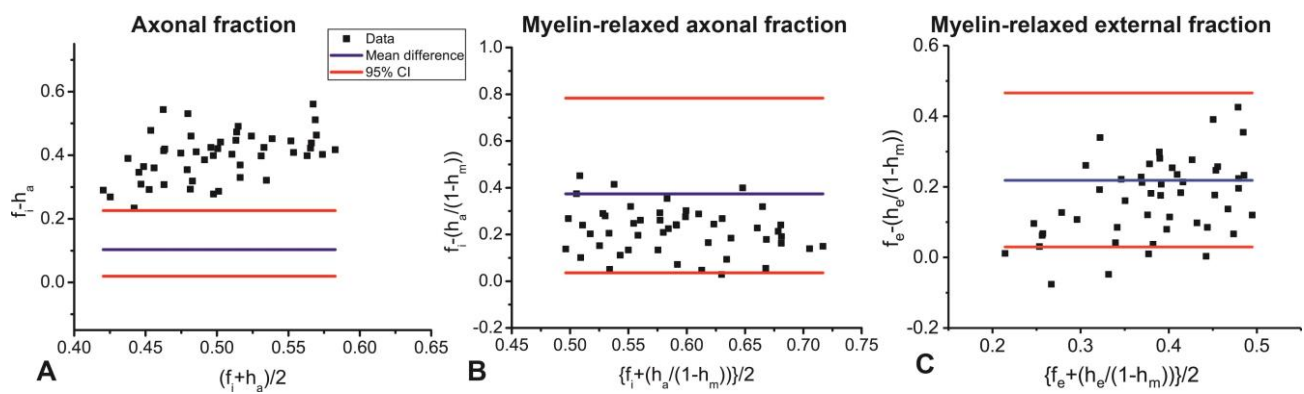

**Figure 7.** Bland-Altman analysis for **(A)** the axonal fraction $h_a$ and **(B)** the myelin-relaxed axonal fraction $h_a/(1-h_m)$ and **(C)** myelin-relaxed external fraction, $h_e/(1-h_m)$, which account for the volume occupied by myelin but which will not be visible in the MRI according to our model (Eq. 1). While $h_a$ is not within the confidence intervals, the myelin-relaxed comparison shows that most of the data resides between the 95% confidence intervals. Therefore, there is a coherent bias in the measurement, but the sought-after quantity – the axonal fraction – is still well-reflected in the data. Around 10% of myelin-relaxed external fraction are outside the 95% confidence interval, which suggest only a moderate agreement.

Inspection of the maps shown in Figure 4 is suggestive of a more complex relationship between the MGE data and the underlying microstructures – and in particular, the regional average axon size seems to be manifest in the data in a non-trivial way. For example, the dCST axons, which are smallest, are also characterized by a relatively large frequency shift, and a rather short $T_{2e}^*$. In lieu of a comprehensive biophysical model describing the complex relationships between the MGE signal decay and the physiology of the tissue, a simple GLM analysis was performed using the ROI data described in Figure 3 and Table 1, and the IHC data given in Tables 2-3 and Figure 5. In particular, we sought to map spatial distribution of average axon diameter using a linear combination of all four MGE-driven parameters from all spinal cords in this study. Figure 8 shows maps of these modeled average axonal sizes, using the following GLM-derived extracted coefficients, $c_0 = 1.3374 \pm 2.6183$; $c_1 = -3.5161 \pm 2.0092$; $c_3 = -0.0096 \pm 0.094$; $c_4 = 0.4963 \pm 0.1441$. Despite that most coefficient errors are quite large, the contrasts arising from the spinal cords when executing a pixel-by-pixel analysis of the GLM provides exquisite contrast which indeed appears to follow the histologically-derived average axonal diameters. Good contrast between the different tracts is evident (Fig. 8), and the smoothness of the transitions between regions, as well as the relatively consistent results across SCs, are noteworthy.

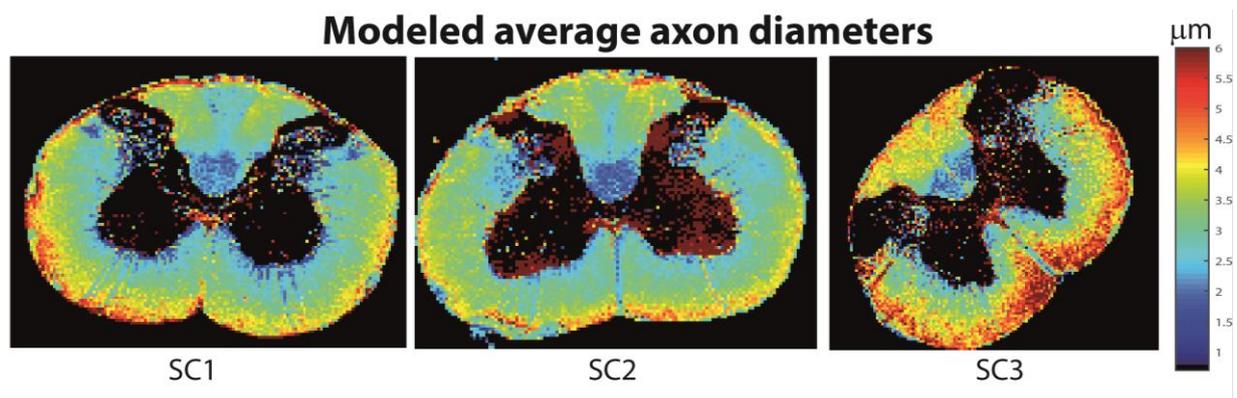

**Figure 8.** Generating the pixel-by-pixel average axonal diameter from the GLM coefficients using all four maps of three representative spinal cords. The maps generated are faithful to the histology (compare with Table 3 for the various ROIs), with smooth transitions across regions, suggesting that the model parameters are consistent across all WM regions.

Finally, it is worth reflecting on whether the "diffraction" patterns in the MGE data could be observed in lower fields. Given lower $T_2$s at lower fields (and of course the different $B_0$ entering the term $\Delta\chi B_0$), we simulated the ensuing signals in Figure 9. The non-monotonicity is expected to be preserved when achieving longer TEs for both these fields.

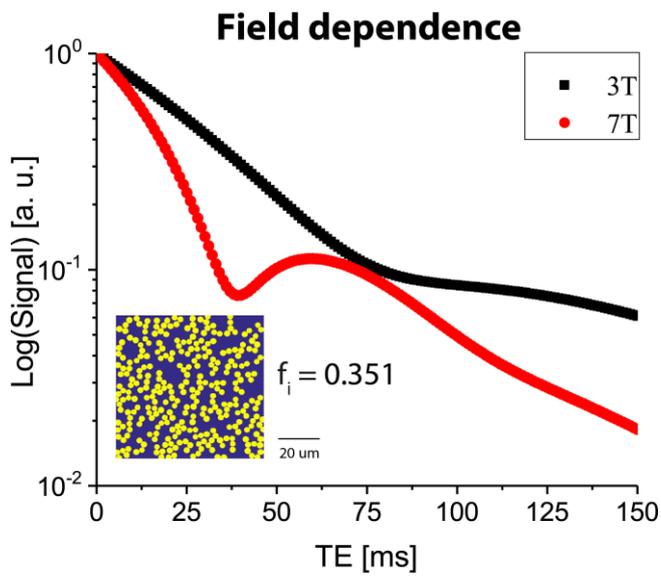

**Figure 9.** Magnetic field dependence. The simulations here show the potential of the MGE approach for 3 and 7T revealing clear non-monotonicity for all fields. Note that different T$_2$ values were assumed, in particular, 85 and 30 ms for T$_{2i}$ and T$_{2e}$ for 3T and 50 and 20 ms for 7Tm respectively.

## Discussion

Imaging regional axonal microstructural properties noninvasively is imperative for longitudinally following and assessing the CNS both under normal conditions, as well as in disease. This is of particular importance when considering the relationships between the axonal diameters and myelination and the conduction velocities (2,3,5,7,9,79), as well as the severe functional deficits arising from aberrations in white matter microstructures, including axon density and regional average diameters (10,16,18,80,81). Diffusion-based MRI methods, such as QSI (19,20,24-26) and temporal diffusion spectroscopy (31,82-84) have been able to provide insights into the regional average axon diameters and density in healthy and diseased CNS; however, these methods typically rely on special hardware or on very long acquisition times, respectively.

One important microstructural property – namely the orientation of the WM fibers – has been quite extensively studied in the context of gradient-echo MRI, both in terms of the phase dependence (52-54,58,62-64,85-87), as well as in terms of $T_2^*$ dependence (59,67-72). Recently, van Gelderen et al have demonstrated non-exponential behavior of the signal decay in humans at both 3 and 7 T, suggesting that the signal may be influenced by multiple compartments (75); Chen et al have simulated signal decays for numerous scenarios, some of which apparently reveal a zero crossings in phase, which – in some cases – could also translate to signal non-monotonicity in the magnitude data (76). Still, these studies focused on distinguishing different populations – mainly myelin vs. axonal water – rather than attempting to experimentally extract the microstructural parameters sought after in this study: axonal density and average diameter.

Our simulations for non-monotonic decay are in accordance with several results given previously, mainly with van Gelderen (75) and with Chen (76), both which seem to predict a signal non-monotonicity as TEs are increased in MGE sequences. Furthermore, a similar multi-compartment model was proposed by others (73,88), but signal non-monotonicities were not observed, presumably due to not reaching sufficiently high echo times for the lower fields studied. Our simulations show that, for the case of fibers aligned perfectly with the main magnetic field, the non-monotonic MGE signals are capable of reflecting the intra-cylindrical fraction independently of the exact cylinder diameter. Our MGE experiments, conducted at 16.4 T on spinal cords aligned with the main magnetic field, have confirmed the existence of the signal non-monotonicity in most WM regions (Figure 3), or, when absent, a non-exponential dependence on TE, in agreement with (75). Importantly, the signal characteristics as modeled by Eq. 1 generate highly informative maps, capable of qualitatively contrasting the major WM tracts and providing several quantitative insights into its microstructure. Our histological findings show that the MRI-extracted intra-cylindrical fraction corresponds to the myelin-relaxed axonal fraction derived from histology, $h_a/(1-h_m)$, but with a systematic shift (Figure 6). Since the myelin signal is "invisible" Eq. 1 appears to describe the data sufficiently well. The correlation with the myelin-relaxed axon fraction showed a small bias which could be detected in the Bland-Altman plot – although clearly the results are all within the 95% confidence limit once the bias is removed (Figure 8). Another histologically-validated finding of this study is the extra-cylindrical fraction, which, by reciprocation, also provides an estimate of the combined axonal + myelin fraction. Furthermore, the fits to the GLM – which were incentivized by the (qualitative) observation that most model parameters depend to some extent on the underlying microstructure (Figure 4) – indicate that the mean axon diameter can be mapped from these simple MGE experiments (Figure 8). Together, these findings suggest a novel way of characterizing salient microstructural features of white matter microstructure using the non-monotonic behavior of a simple MGE experiment.

Our choice of 2 compartment model may be oversimplified, and a clear alternative would be a 3 compartment model that takes into account myelin water. Preliminary tests with the 3 compartment model showed no evidence for better correlations with the data, and in fact the fits tended to converge to two similar components and another distinct one, suggesting that - at least at this high field of 16.4 T – the 2 compartment model here used was sufficient. More theory and experiments are required for assessing the contribution of myelin. Differences between ex- and in-vivo tissues are also important to investigate as their susceptibility may vary (see, e.g., Ref. 64). Nevertheless, the excellent contrasts here observed provide a good starting point for future studies.

In contradistinction with our simulations and with the theory, we observed significant experimental deviations from our simple model. In particular, the unexpected region-specific frequency shift was observed in the spinal cords (Figure 4). Given that in our simple model, the frequency shift should only depend on the quantity $\Delta\omega = \gamma\Delta\chi B_0$, and that according to the model this quantity is fixed – it is not expected to vary much across areas characterized by different axon density or size. Our findings of a non-constant frequency shift across the different ROIs (Table 1), as well as its qualitative dependence on axon diameter (Figure 4), likely may reflect several more complex aspects of the microstructure, which require more comprehensive theoretical treatment to account for. Several factors may potentially play a role in the observed frequency dependence: (1) the existence of more than two compartments: in particular, myelin water may experience stronger frequency shifts compared with axonal water (73,76); however, histologically-derived myelin fractions were found to be relatively constant across areas (Table 2), and at 16.4 T it is plausible that myelin relaxation times are fairly short, thereby reducing the likelihood of variations in myelin to be the most prominent source of variation in frequency shift across the different WM tracts in the spinal cord; (2) misalignment of the spinal cord or axonal dispersion: deviations from a perfect alignment with respect to $B_0$ may induce a more complex pattern of spatially-varying magnetic field distributions, giving rise to slightly different shifts in the different tracts. However, given that these effects are (again, ideally) proportional to $B_0 \cos^2(\theta)$, where $\theta$ is the polar angle between the cylinder direction and magnetic field, and given the excellent alignment observed in the DTI analysis (Figure 2), we expect these effects to be negligible as $\theta$ is very small; (3) diffusion and exchange effects: given that axons are not perfectly aligned nor cylindrical, a variation in the magnetic field around the axon is plausible. These internal, susceptibility-induced magnetic field gradients may be experienced by diffusing spins, which may then contribute to a smaller frequency shift in those compartments (motion narrowing), thereby leading to an increase in the susceptibility difference between the compartments. Indeed, our experiments do not provide conclusive evidence to answer whether the spins here observed are in the static dephasing regime or in the motional averaging regime.

An interesting question given that the data seems to be fairly well explained by two compartments - is the actual assignment of the two compartments. In our simple model, we suggested that a susceptibility difference exists between the intra/extra-cylindrical spaces, while ignoring the myelin. Clearly, identical signal decays could be obtained by assuming that the intra/extra-cylindrical susceptibilities are equal (i.e., they represent a single component), while the second component would reside in the myelin (and assuming that the volume fractions add up similarly). Our experiments cannot at this point determine which of these two scenarios is the correct one, and we feel it is still premature to assign these compartments to specific tissue components based on the experiments provided herein. More simulations, experiments including, e.g., magnetic field dependence of the signal decay or filtering the data using other relaxation mechanisms, as well as advances to the biophysical model, are required to pinpoint the origin of the susceptibility-driven oscillations reported here. Magnetic Field Correlations (89,90), combined with the GLTA approaches (64) may serve as a good starting point.

This study focused on the case of parallel white matter, and did not directly consider other orientations. This served to simplify an otherwise rather complex scenario, whereby, more generally, each orientation would generate a unique spatially dependent frequency distribution, which would also depend on its radius. This also obviates potential confounds arising from diffusion in these density- and size-dependent field distributions. In the particular case of spinal cord, it may also correspond more closely to the practical situation in clinical settings, where at least some of the spinal cord is fairly parallel to the field in the supine position. Again, more simulations and experiments are required to test whether the axon density-dependent signal non-monotonicity persists for a wide range of polar angles and densities. As well, the effects of axonal diameter distributions which was not accounted for in this study, may have effects on the form of the MGE signal decay.

Our experiments were conducted at 16.4 T, where the frequency shift term was found to be around ~25-50 Hz. At lower fields, which are much more common both in preclinical and clinical research, these microstructural frequency shifts are expected to be smaller, and, depending on the respective $T_2^*$s of the different compartments, the oscillations observed in our signals may be shifted to higher TE values. Another consideration is the background field inhomogeneity, whose source is mainly imperfect shimming, and whose effects on the signal decay were neglected in Eq. 1. A background field distortion would incur an additional source of frequency shift (70,73,75,91); however, it is expected to be smooth across the sample. In our study, despite the high field, the typical linewidths across the entire sample were relatively narrow, ~20-30 Hz after shimming. In addition, we performed similar MGE experiments but with much higher resolution (10x10x250 (μm)$^3$) and found similar contrasts (data not shown); as the background field contribution is strongly dependent on voxel size (lower for decreasing voxel size), we conclude that the magnetic field inhomogeneity seems to have contributed only little to the experimental results given in this study. In the future, filters designed to remove the background field variation, such as those typically employed in quantitative susceptibility mapping (92), could be used to mitigate such putative deleterious effects.

From a methodological perspective, it is worth noting that the MGE is arguably one of MRI's simplest sequences, providing very high SNR per unit time due to its ability to execute efficient Ernst angle acquisition schemes. In addition, MGE does not require multiple shots to record the entire TE decay, since it simply uses gradient refocusing and re-readout of the same k-space line following a single excitation, to achieve the entire TE-dependence (per k-space line) in a single shot. Finally, MGE is SAR efficient, requiring no refocusing pulses, which, at high field, may be potentially confounding for spin-echo-based sequence. Clearly, gaining direct and noninvasive access to the histological internal water fraction ($h_{int} = f_a + f_m$) and the extra-axonal fraction ($h_e$), and a somewhat more indirect access to the average axon size and axon density, as here shown, using such a simple MGE sequence, could offer a real advantage in microstructural imaging. More studies are required to assess the potential for these types of sequences in the clinic and in lower fields, and to directly compare diffusion-based methods with the MGE approach. However, the first results shown here, correlating the MGE findings to histology, suggest a potential role for MGE in microstructural imaging *in-vivo*.

## Conclusions

The potential of MGE experiments to map axonal microstructures in the spinal cord were investigated using simulations, experiments at ultrahigh field, and correlations with histology. For parallel white matter, MGE experiments executed up to high TEs reveal high contrast in microstructurally distinct regions. Signal oscillations were detected, from which revealing parametric maps were extracted using a simple two-compartment model. Correlations against ground-truth histological findings suggest that the extracted density maps reflect axonal, internal, and extra-axonal fractions, and that average axon diameters can be mapped by empirically utilizing the four model parameters. These features are promising for future high-fidelity microstructural characterizations of both healthy and diseased CNS.

**Acknowledgements.**

NS gratefully acknowledges funding from the European Research Council (ERC) under the European Union's Horizon 2020 research and innovation programme (grant agreement No. 679058 - DIRECT-fMRI) As well as under the Marie Sklodowska-Curie grant agreement No 657366.

# Figure Captions

**Figure 1.** Simulations for MGE signal decays in perfectly aligned cylinders. (A) Frequency distribution maps for the microstructures considered here. Cylinder density was varied from ~0.1-0.4. Notice that two discrete frequencies emerge in these plots (blue and yellow). (B) Echo-time dependence of the magnitude of the signal, |S(TE)|. Notice the density-dependent signal behaviour, including the non-monotonic, diffraction-like behaviour. $f_i$ denotes the intra-cylindrical fraction. (C) Further simulations conducted for the narrower range $f_i$ = 0.25-0.35, showing that densities have quite unique signatures for the microstructure in these MGE magnitude data.

**Figure 2.** Alignment of the spinal cord with respect to B0 in a representative rat spinal cord. Color-coded FA maps extracted from the DTI scan reveal predominantly parallel orientation in white matter (blue denotes the magnet's z-direction, the direction of the field). Red color reflects left-right orientation and green up-down. The histogram of angles in white matter shows that the majority of white matter voxels are aligned within ~3° of the z-axis.

**Figure 3.** ROI analysis in three representative spinal cords. ROIs were placed within the seven major tracts running along the spinal cord (left column). The average signal in the ROI is plotted in the middle column, showing the signal decay (symbols) and the fits (solid lines) to the data using Eq. 1. A consistent non-monotonicity was observed for numerous ROIs, in a consistent fashion across specimens. The rightmost column expands the TE=0-25 ms region, to better highlight the oscillations and fits. The signals are plotted in log scale for easier viewing of the oscillation. dCST – dorsal corticospinal tract; FG – Fasiculus Gracilis; FC – Fasiculus Cuneatis; ReST – Reticulospinal tract; RST – Rubrospinal tract; STT – Spinothalamic tract; VST – Vestibulospinal tract.

**Figure 4.** Parametric maps arising from pixel-by-pixel fits of the TE-dependence for three representative spinal cords. These maps revealed highly consistent internal fractions $f_i$, relaxation constants, and frequency shifts (SC3 represents the lower-most observed values in the entire study). A pronounced contrast between the major spinal cord tracts is observed in these maps, suggesting their correlation with underlying microstructural features.

**Figure 5.** Histological study. **(A)** Wide field images of the three stains used in this study, designed to capture cell nuclei (DAPI, blue), myelin (PLP, green), and axons (NF, red). **(B)** Multichannel analysis of the different stains in different regions. Note the overlay contains no overlapping signals, showing that the antibodies were indeed selective for their target. Scale bar, 50 µm. **(C)** A zoomed view of (B) for the seven different regions of interest in the spinal cord white matter. These data were used for the quantification of all histologically-derived parameters.

**Figure 6.** Correlation between MRI-derived parameters and IHC-derived parameters. The axonal fraction contains only intra-axonal space; the internal fraction contains both axonal and myelin spaces; and the extra-axonal water contains no water from the internal fraction. Scatter plots of the MRI-driven intra-cylindrical fraction $f_i$ against the IHC-derived **(A)** axonal fraction $h_a$, **(B)**, myelin-relaxed axonal fraction $h_a/(1-h_m)$, **(C)** internal fraction $h_{int}=h_a+h_m$. **(D)** MRI-extracted extra-cylindrical fraction compared against the extra-axonal fraction derived from IHC, **(E)** myelin-relaxed axonal fraction $h_e/(1-h_m)$. Note that the MRI-derived $f_i$ correlates well with the IHC-derived internal fraction $h_{int}$ but not with the axonal fraction $h_a$. The MRI-derived $f_e = 1-f_i$ correlates well with the histological extra-axonal fraction.

**Figure 7.** Bland-Altman analysis for **(A)** the axonal fraction $h_a$ and **(B)** the myelin-relaxed axonal fraction $h_a/(1-h_m)$ and **(C)** myelin-relaxed external fraction, $h_e/(1-h_m)$, which account for the volume occupied by myelin but which will not be visible in the MRI according to our model (Eq. 1). While $h_a$ is not within the confidence intervals, the myelin-relaxed comparison shows that most of the data resides between the 95% confidence intervals. Therefore, there is a coherent bias in the measurement, but the sought-after quantity – the axonal fraction – is still well-reflected in the data. Around 10% of myelin-relaxed external fraction are outside the 95% confidence interval, which suggest only a moderate agreement.

**Figure 8.** Generating the pixel-by-pixel average axonal diameter from the GLM coefficients using all four maps of three representative spinal cords. The maps generated are faithful to the histology (compare with Table 3 for the various ROIs), with smooth transitions across regions, suggesting that the model parameters are consistent across all WM regions.

**Figure 9.** Magnetic field dependence. The simulations here show the potential of the MGE approach for 3 and 7T revealing clear non-monotonicity for all fields. Note that different $T_2$ values were assumed, in particular, 85 and 30 ms for $T_{2i}$ and $T_{2e}$ for 3T and 50 and 20 ms for 7T respectively.

# References

1. Kandel ER, Schwartz JH, and Jessell TM. Principles of Neural Sciences 4th ed. Mcgraw-Hill 2000.
2. Caminiti R, Carducci F, Piervincenzi C, Battaglia-Mayer A, Confalone G, Visco-Comandini F, Pantano P, and Innocenti GM. Diameter, Length, Speed, and Conduction Delay of Callosal Axons in Macaque Monkeys and Humans: Comparing Data from Histology and Magnetic Resonance Imaging Diffusion Tractography. Journal of Neuroscience 2013;33:14501-14511.
3. Cullheim S. Relations Between Cell Body Size, Axon Diameter and Axon Conduction-Velocity of Cat Sciatic Alpha-Motoneurons Stained with Horseradish-Peroxidase. Neuroscience Letters 1978;8:17-20.
4. Innocenti GM, Vercelli A, and Caminiti R. The Diameter of Cortical Axons Depends Both on the Area of Origin and Target. Cereb Cortex 2013.
5. Perge JA, Niven JE, Mugnaini E, Balasubramanian V, and Sterling P. Why Do Axons Differ in Caliber? Journal of Neuroscience 2012;32:626-638.
6. Sherman DL and Brophy PJ. Mechanisms of axon ensheathment and myelin growth. Nat Rev Neurosci 2005;6:683-690.
7. Caminiti R, Ghaziri H, Galuske R, Hof PR, and Innocenti GM. Evolution amplified processing with temporally dispersed slow neuronal connectivity in primates. Proceedings of the National Academy of Sciences of the United States of America 2009;106:19551-19556.
8. Innocenti GM. Development and evolution: Two determinants of cortical connectivity. Gene Expression to Neurobiology and Behavior: Human Brain Development and Developmental Disorders 2011;189:65-75.
9. Tomasi S, Caminiti R, and Innocenti GM. Areal Differences in Diameter and Length of Corticofugal Projections. Cerebral Cortex 2012;22:1463-1472.
10. Payne SC, Bartlett CA, Harvey AR, Dunlop SA, and Fitzgerald M. Myelin sheath decompaction, axon swelling, and functional loss during chronic secondary degeneration in rat optic nerve. Invest Ophthalmol Vis Sci 2012;53:6093-6101.
11. Payne SC, Bartlett CA, Harvey AR, Dunlop SA, and Fitzgerald M. Chronic swelling and abnormal myelination during secondary degeneration after partial injury to a central nervous system tract. J Neurotrauma 2011;28:1077-1088.
12. Deluca GC, Ebers GC, and Esiri MM. Axonal loss in multiple sclerosis: a pathological survey of the corticospinal and sensory tracts. Brain 2004;127:1009-1018.
13. Evangelou N, Konz D, Esiri MM, Smith S, Palace J, and Matthews PM. Size-selective neuronal changes in the anterior optic pathways suggest a differential susceptibility to injury in multiple sclerosis. Brain 2001;124:1813-1820.
14. Lovas G, Szilagyi N, Majtenyi K, Palkovits M, and Komoly S. Axonal changes in chronic demyelinated cervical spinal cord plaques. Brain 2000;123:308-317.
15. Maxwell WL, Bartlett E, and Morgan H. Wallerian degeneration in the optic nerve stretch-injury model of traumatic brain injury: a stereological analysis. J Neurotrauma 2015;32:780-790.
16. Nashmi R and Fehlings MG. Changes in axonal physiology and morphology after chronic compressive injury of the rat thoracic spinal cord. Neuroscience 2001;104:235-251.
17. Budde MD and Frank JA. Neurite beading is sufficient to decrease the apparent diffusion coefficient after ischemic stroke. Proceedings of the National Academy of Sciences of the United States of America 2010;107:14472-14477.
18. Lai WS, Xu B, Westphal KGC, Paterlini M, Olivier B, Pavlidis P, Karayiorgou M, and Gogos JA. Akt1 deficiency affects neuronal morphology and predisposes to abnormalities in prefrontal cortex functioning. Proceedings of the National Academy of Sciences of the United States of America 2006;103:16906-16911.
19. Assaf Y, Mayk A, and Cohen Y. Displacement imaging of spinal cord using q-space diffusion-weighted MRI. Magn Reson Med 2000;44:713-722.
20. Assaf Y, Ben-Bashat D, Chapman J, Peled S, Biton IE, Kafri M, Segev Y, Hendler T, Korczyn AD, Graif M, and Cohen Y. High b-value q-space analyzed diffusion-weighted MRI: application to multiple sclerosis. Magn Reson Med 2002;47:115-126.
21. Biton IE, Mayk A, Kidron D, Assaf Y, and Cohen Y. Improved detectability of experimental allergic encephalomyelitis in excised swine spinal cords by high b-value q-space DWI. Exp Neurol 2005;195:437-446.

22. Cohen Y and Assaf Y. High b-value q-space analyzed diffusion-weighted MRS and MRI in neuronal tissues - a technical review. NMR in Biomedicine 2002;15:516-542.
23. Farrell JA, Smith SA, Gordon-Lipkin EM, Reich DS, Calabresi PA, and van Zijl PC. High b-value q-space diffusion-weighted MRI of the human cervical spinal cord in vivo: feasibility and application to multiple sclerosis. Magn Reson Med 2008;59:1079-1089.
24. Nossin-Manor R, Duvdevani R, and Cohen Y. Spatial and temporal damage evolution after hemi-crush injury in rat spinal cord obtained by high b-value q-space diffusion magnetic resonance imaging. J Neurotrauma 2007;24:481-491.
25. Ong HH, Wright AC, Wehrli SL, Souza A, Schwartz ED, Hwang SN, and Wehrli FW. Indirect measurement of regional axon diameter in excised mouse spinal cord with q-space imaging: Simulation and experimental studies. Neuroimage 2008;40:1619-1632.
26. Ong HH and Wehrli FW. Quantifying axon diameter and intra-cellular volume fraction in excised mouse spinal cord with q-space imaging. Neuroimage 2010;51:1360-1366.
27. Aboitiz F, Scheibel AB, Fisher RS, and Zaidel E. Fiber Composition of the Human Corpus-Callosum. Brain Research 1992;598:143-153.
28. Barazany D, Basser PJ, and Assaf Y. In vivo measurement of axon diameter distribution in the corpus callosum of rat brain. Brain 2009;132:1210-1220.
29. Innocenti GM, Caminiti R, and Aboitiz F. Comments on the paper by Horowitz et al. (2014). Brain Struct Funct 2015;220:1789-1790.
30. Duval T, McNab JA, Setsompop K, Witzel T, Schneider T, Huang SY, Keil B, Klawiter EC, Wald LL, and Cohen-Adad J. In vivo mapping of human spinal cord microstructure at 300mT/m. Neuroimage 2015;118:494-507.
31. Xu J, Li H, Harkins KD, Jiang X, Xie J, Kang H, Does MD, and Gore JC. Mapping mean axon diameter and axonal volume fraction by MRI using temporal diffusion spectroscopy. Neuroimage 2014;103:10-19.
32. Alvarez GA, Shemesh N, and Frydman L. Coherent dynamical recoupling of diffusion-driven decoherence in Magnetic Resonance. Physical Review Letters 2013;111:080404.
33. Shemesh N, Alvarez GA, and Frydman L. Size Distribution Imaging by Non-Uniform Oscillating-Gradient Spin Echo (NOGSE) MRI. PLoS One 2015;10:e0133201.
34. Huang SY, Nummenmaa A, Witzel T, Duval T, Cohen-Adad J, Wald LL, and McNab JA. The impact of gradient strength on in vivo diffusion MRI estimates of axon diameter. Neuroimage 2015;106:464-472.
35. Alexander DC, Hubbard PL, Hall MG, Moore EA, Ptito M, Parker GJM, and Dyrby TB. Orientationally invariant indices of axon diameter and density from diffusion MRI. Neuroimage 2010;52:1374-1389.
36. Dyrby TB, Sogaard LV, Hall MG, Ptito M, and Alexander DC. Contrast and stability of the axon diameter index from microstructure imaging with diffusion MRI. Magn Reson Med 2013;70:711-721.
37. Ianus A, Drobnjak I, and Alexander DC. Model-based estimation of microscopic anisotropy using diffusion MRI: a simulation study. NMR Biomed 2016;29:672-685.
38. Sepehrband F, Alexander DC, Clark KA, Kurniawan ND, Yang Z, and Reutens DC. Parametric Probability Distribution Functions for Axon Diameters of Corpus Callosum. Front Neuroanat 2016;10:59.
39. Zhang H, Hubbard PL, Parker GJ, and Alexander DC. Axon diameter mapping in the presence of orientation dispersion with diffusion MRI. Neuroimage 2011;56:1301-1315.
40. Zhang H, Dyrby TB, and Alexander DC. Axon diameter mapping in crossing fibers with diffusion MRI. Med Image Comput Comput Assist Interv 2011;14:82-89.
41. Zhang H, Schneider T, Wheeler-Kingshott CA, and Alexander DC. NODDI: practical in vivo neurite orientation dispersion and density imaging of the human brain. Neuroimage 2012;61:1000-1016.
42. Jespersen SN, Kroenke CD, Ostergaard L, Ackerman JJ, and Yablonskiy DA. Modeling dendrite density from magnetic resonance diffusion measurements. Neuroimage 2007;34:1473-1486.
43. Jespersen SN, Bjarkam CR, Nyengaard JR, Chakravarty MM, Hansen B, Vosegaard T, Ostergaard L, Yablonskiy D, Nielsen NC, and Vestergaard-Poulsen P. Neurite density from magnetic resonance diffusion measurements at ultrahigh field: Comparison with light microscopy and electron microscopy. Neuroimage 2010;49:205-216.
44. Finsterbusch J. Multiple-Wave-Vector Diffusion-Weighted NMR. Annual Reports on Nmr Spectroscopy, Vol 72 2010;72:225-299.

45. Koch MA and Finsterbusch J. Compartment size estimation with double wave vector diffusion-weighted Imaging. Magnetic Resonance in Medicine 2008;60:90-101.
46. Shemesh N and Cohen Y. Microscopic and Compartment Shape Anisotropies in Gray and White Matter Revealed by Angular Bipolar Double-PFG MR. Magnetic Resonance in Medicine 2011;65:1216-1227.
47. Shemesh N, Rosenberg JT, Dumez JN, Muniz JA, Grant SC, and Frydman L. Metabolic properties in stroked rats revealed by relaxation-enhanced magnetic resonance spectroscopy at ultrahigh fields. Nat Commun 2014;5:4958.
48. Szczepankiewicz F, Lasic S, van Westen D, Sundgren PC, Englund E, Westin CF, Stahlberg F, Latt J, Topgaard D, and Nilsson M. Quantification of microscopic diffusion anisotropy disentangles effects of orientation dispersion from microstructure: Applications in healthy volunteers and in brain tumors. Neuroimage 2015;104:241-252.
49. Jespersen SN, Lundell H, Sonderby CK, and Dyrby TB. Orientationally invariant metrics of apparent compartment eccentricity from double pulsed field gradient diffusion experiments. NMR Biomed 2013;26:1647-1662.
50. Haacke EM, Liu S, Buch S, Zheng W, Wu D, and Ye Y. Quantitative susceptibility mapping: current status and future directions. Magnetic Resonance Imaging 2015;33:1-25.
51. Reichenbach JR, Schweser F, Serres B, and Deistung A. Quantitative Susceptibility Mapping: Concepts and Applications. Clin Neuroradiol 2015;25 Suppl 2:225-230.
52. Lee J, Shmueli K, Fukunaga M, van Gelderen P, Merkle H, Silva AC, and Duyn JH. Sensitivity of MRI resonance frequency to the orientation of brain tissue microstructure. Proc Natl Acad Sci U S A 2010;107:5130-5135.
53. Liu C. Susceptibility tensor imaging. Magn Reson Med 2010;63:1471-1477.
54. Liu C and Li W. Imaging neural architecture of the brain based on its multipole magnetic response. Neuroimage 2013;67:193-202.
55. Duyn JH and Barbara TM. Sphere of Lorentz and demagnetization factors in white matter. Magn Reson Med 2014;72:1-3.
56. Duyn JH. Frequency shifts in the myelin water compartment. Magn Reson Med 2014;71:1953-1955.
57. Duyn JH and Schenck J. Contributions to magnetic susceptibility of brain tissue. NMR Biomed 2016.
58. Lee J, Shmueli K, Kang BT, Yao B, Fukunaga M, van Gelderen P, Palumbo S, Bosetti F, Silva AC, and Duyn JH. The contribution of myelin to magnetic susceptibility-weighted contrasts in high-field MRI of the brain. Neuroimage 2012;59:3967-3975.
59. van Gelderen P, Mandelkow H, de Zwart JA, and Duyn JH. A torque balance measurement of anisotropy of the magnetic susceptibility in white matter. Magn Reson Med 2015;74:1388-1396.
60. Wharton S and Bowtell R. Effects of white matter microstructure on phase and susceptibility maps. Magn Reson Med 2015;73:1258-1269.
61. Yablonskiy DA, Luo J, Sukstanskii AL, Iyer A, and Cross AH. Biophysical mechanisms of MRI signal frequency contrast in multiple sclerosis. Proc Natl Acad Sci U S A 2012;109:14212-14217.
62. Yablonskiy DA, He X, Luo J, and Sukstanskii AL. Lorentz sphere versus generalized Lorentzian approach: What would lorentz say about it? Magn Reson Med 2014;72:4-7.
63. Yablonskiy DA and Sukstanskii AL. Biophysical mechanisms of myelin-induced water frequency shifts. Magn Reson Med 2014;71:1956-1958.
64. Yablonskiy DA and Sukstanskii AL. Generalized Lorentzian Tensor Approach (GLTA) as a biophysical background for quantitative susceptibility mapping. Magn Reson Med 2015;73:757-764.
65. He X and Yablonskiy DA. Biophysical mechanisms of phase contrast in gradient echo MRI. Proceedings of the National Academy of Sciences of the United States of America 2009;106:13558-13563.
66. Sukstanskii AL and Yablonskiy DA. On the role of neuronal magnetic susceptibility and structure symmetry on gradient echo MR signal formation. Magn Reson Med 2014;71:345-353.
67. Lee J, van Gelderen P, Kuo LW, Merkle H, Silva AC, and Duyn JH. T2*-based fiber orientation mapping. Neuroimage 2011;57:225-234.
68. Sati P, Silva AC, van Gelderen P, Gaitan MI, Wohler JE, Jacobson S, Duyn JH, and Reich DS. In vivo quantification of T(2) anisotropy in white matter fibers in marmoset monkeys. Neuroimage 2012;59:979-985.

69. Wharton S and Bowtell R. Fiber orientation-dependent white matter contrast in gradient echo MRI. Proc Natl Acad Sci U S A 2012;109:18559-18564.
70. Wharton S and Bowtell R. Gradient echo based fiber orientation mapping using R2* and frequency difference measurements. Neuroimage 2013;83:1011-1023.
71. Cohen-Adad J, Polimeni JR, Helmer KG, Benner T, McNab JA, Wald LL, Rosen BR, and Mainero C. T(2)* mapping and B(0) orientation-dependence at 7 T reveal cyto- and myeloarchitecture organization of the human cortex. Neuroimage 2012;60:1006-1014.
72. Rudko DA, Klassen LM, de Chickera SN, Gati JS, Dekaban GA, and Menon RS. Origins of R2* orientation dependence in gray and white matter. Proc Natl Acad Sci U S A 2014;111:E159-E167.
73. Sati P, van Gelderen P, Silva AC, Reich DS, Merkle H, de Zwart JA, and Duyn JH. Micro-compartment specific T2* relaxation in the brain. Neuroimage 2013;77:268-278.
74. Sukstanskii AL, Wen J, Cross AH, and Yablonskiy DA. Simultaneous multi-angular relaxometry of tissue with MRI (SMART MRI): Theoretical background and proof of concept. Magn Reson Med 2016.
75. van Gelderen P, de Zwart JA, Lee J, Sati P, Reich DS, and Duyn JH. Nonexponential T(2) decay in white matter. Magn Reson Med 2012;67:110-117.
76. Chen WC, Foxley S, and Miller KL. Detecting microstructural properties of white matter based on compartmentalization of magnetic susceptibility. Neuroimage 2013;70:1-9.
77. Storey P, Chung S, Ben-Eliezer N, Lemberskiy G, Lui YW, and Novikov DS. Signatures of microstructure in conventional gradient and spin echo signals. Proc Intl Soc Mag Reson Med 23 (2015) 2014:0014.
78. Nunes D and Kuner T. Disinhibition of olfactory bulb granule cells accelerates odour discrimination in mice. Nat Commun 2015;6.
79. Cole JS, Messing A, Trojanowski JQ, and Lee VM. Modulation of axon diameter and neurofilaments by hypomyelinating Schwann cells in transgenic mice. J Neurosci 1994;14:6956-6966.
80. Nikic I, Merkler D, Sorbara C, Brinkoetter M, Kreutzfeldt M, Bareyre FM, Bruck W, Bishop D, Misgeld T, and Kerschensteiner M. A reversible form of axon damage in experimental autoimmune encephalomyelitis and multiple sclerosis. Nat Med 2011;17:495-499.
81. Cusack CL, Swahari V, Henley WH, Ramsey JM, and Deshmukh M. Distinct pathways mediate axon degeneration during apoptosis and axon-specific pruning. Nature Communications 2013;4.
82. Xu J, Does MD, and Gore JC. Quantitative characterization of tissue microstructure with temporal diffusion spectroscopy. J Magn Reson 2009;200:189-197.
83. Gore JC, Xu JZ, Colvin DC, Yankeelov TE, Parsons EC, and Does MD. Characterization of tissue structure at varying length scales using temporal diffusion spectroscopy. NMR in Biomedicine 2010;23:745-756.
84. Does MD, Parsons EC, and Gore JC. Oscillating gradient measurements of water diffusion in normal and globally ischemic rat brain. Magnetic Resonance in Medicine 2003;49:206-215.
85. Li W, Liu C, Duong TQ, van Zijl PC, and Li X. Susceptibility tensor imaging (STI) of the brain. NMR Biomed 2016.
86. Liu C, Li W, Johnson GA, and Wu B. High-field (9.4 T) MRI of brain dysmyelination by quantitative mapping of magnetic susceptibility. Neuroimage 2011;56:930-938.
87. Liu C, Li W, Wu B, Jiang Y, and Johnson GA. 3D fiber tractography with susceptibility tensor imaging. Neuroimage 2012;59:1290-1298.
88. Nam Y, Lee J, Hwang D, and Kim DH. Improved estimation of myelin water fraction using complex model fitting. Neuroimage 2015;116:214-221.
89. Jensen JH, Chandra R, Ramani A, Lu H, Johnson G, Lee SP, Kaczynski K, and Helpern JA. Magnetic field correlation imaging. Magn Reson Med 2006;55:1350-1361.
90. Jensen JH, Szulc K, Hu C, Ramani A, Lu H, Xuan L, Falangola MF, Chandra R, Knopp EA, Schenck J, Zimmerman EA, and Helpern JA. Magnetic field correlation as a measure of iron-generated magnetic field inhomogeneities in the brain. Magn Reson Med 2009;61:481-485.
91. Shmueli K, de Zwart JA, van Gelderen P, Li TQ, Dodd SJ, and Duyn JH. Magnetic susceptibility mapping of brain tissue in vivo using MRI phase data. Magn Reson Med 2009;62:1510-1522.
92. Schweser F, Deistung A, Sommer K, and Reichenbach JR. Toward online reconstruction of quantitative susceptibility maps: superfast dipole inversion. Magn Reson Med 2013;69:1582-1594.